\documentclass[10pt,conference]{IEEEtran}
\IEEEoverridecommandlockouts
\usepackage{color}
\usepackage{cite}
\usepackage{verbatim}
\usepackage{bm}
\usepackage{amsmath,amssymb,amsfonts}
\usepackage[bottom=1in,top=1.8cm,left=1.7cm,right=1.7cm]{geometry}
\usepackage{algorithmic}
\usepackage{algorithm}
\usepackage{graphicx}
\usepackage{textcomp}
\usepackage{xcolor}
\usepackage{graphicx}
\usepackage{subfigure}
\def\BibTeX{{\rm B\kern-.05em{\sc i\kern-.025em b}\kern-.08em
    T\kern-.1667em\lower.7ex\hbox{E}\kern-.125emX}}
    
\IEEEoverridecommandlockouts\IEEEpubid{\makebox[\columnwidth]{ 978-1-6654-3540-6/22~\copyright~2022 IEEE \hfill} \hspace{\columnsep}\makebox[\columnwidth]{ }}

\begin{document}
\title{Geo-Spatio-Temporal Information Based 3D Cooperative Positioning in LOS/NLOS Mixed Environments\\
\thanks{
This work was supported in part by Beijing Municipal Natural Science Foundation under Grant L202012 and Grant Z220004, in part by the National Key Research and Development Program of China under Grant 2020YFA0711302, and in part by the Fundamental Research Funds for the Central Universities under Grant 2020RC05. (\textit{Corresponding
author: Shaoshi Yang})
}
}

\author{Yue~Cao,
        Shaoshi~Yang,~\IEEEmembership{Senior Member,~IEEE},
        and Zhiyong~Feng,~\IEEEmembership{Senior Member,~IEEE}
        \thanks{The authors are with the Key Laboratory of Universal Wireless Communications, Ministry of Education, and the School of Information and Communication Engineering, Beijing University of Posts and Telecommunications, Beijing, 100876, 
        China (e-mails: \{caoyue, shaoshi.yang, fengzy\}@bupt.edu.cn).
        
        978-1-6654-3540-6/22/ \$31.00 ©2022 IEEE
        }
        }

\markboth{}%
{Shell \MakeLowercase{\textit{et al.}}: Bare Demo of IEEEtran.cls for Computer Society Journals}

\maketitle
\pagestyle{empty}  
\thispagestyle{empty} 

\begin{abstract}
We propose a geographic and spatio-temporal information based distributed cooperative positioning (GSTICP) algorithm for wireless networks that require three-dimensional (3D) coordinates and operate in the line-of-sight (LOS) and non-line-of-sight (NLOS) mixed environments. First, a factor graph (FG) is created by factorizing the \textit{a posteriori} distribution of the position-vector estimates and mapping the spatial-domain and temporal-domain operations of nodes onto the FG. Then, we exploit a geographic information based NLOS identification scheme to reduce the performance degradation caused by NLOS measurements. Furthermore, we utilize a finite symmetric sampling based scaled unscented transform (SUT) method to approximate the nonlinear terms of the messages passing on the FG with high precision, despite using only a small number of samples. Finally, we propose an enhanced anchor upgrading (EAU) mechanism to avoid redundant iterations. Our GSTICP algorithm supports any type of ranging measurement that can determine the distance between nodes. Simulation results and analysis demonstrate that our GSTICP has a lower computational complexity than the state-of-the-art belief propagation (BP) based localizers, while achieving an even more competitive positioning performance.
\end{abstract}

\begin{IEEEkeywords} Cooperative positioning, factor graph, spatio-temporal information, scaled unscented transform, network localization, Internet of Things (IoT).
\end{IEEEkeywords}

\section{Introduction}
Location awareness is crucial in many emerging civilian and military applications \cite{b0, b1,b12,b13} relying on wireless networks, e.g., wireless sensor network (WSN) and the Internet of Things (IoT). Distributed cooperative positioning (CP) \cite{b2,b3} is a promising technology capable of providing the reliable location information for wireless networks operating in global navigation satellite system (GNSS) denied environments, such as dense urban areas and underground spaces.

Distributed CP algorithms are usually based on probabilistic models, which consider all parameters to be random variables and make each node to be localized (usually called \textit{agent}) infer its own position based on its locally available information, including the external measurements and internal measurements (if exist). In order to better model a distributed wireless network using its locality structure and better exploit the time correlation between past and current epochs, a class of distributed CP algorithms based on the factor graph (FG) framework were proposed in \cite{b4, b5, b6}, where FG enables solving the marginal function of multivariate global functions more efficiently and is more suitable for distributed implementation than traditional belief propagation (BP). However, the approaches in \cite{b4} and \cite{b6} both utilized massive sample points to approximate the nonlinear terms of the iterative messages passing on the FG, thus resulting in high computational complexity and communication overhead. Additionally, the scheme of \cite{b5} achieved a low computational complexity and communication overhead, but it sacrificed the positioning accuracy by employing the first order Taylor expansion (TE) to replace nonlinear terms in the messages of the parametric sum-product algorithm (SPA).

Unfortunately, the lack of appropriate methods to process the non-line-of-sight (NLOS) measurements prevents most existing CP algorithms from being applied to practical wireless networks. The positive bias caused by NLOS measurements may propagate throughout the whole wireless network and degrade the positioning accuracy dramatically. Attempts have been made to identify the NLOS measurements by using the eigenvalue of the error matrix which contains the noise and NLOS bias \cite{b7}. In \cite{b8}, the authors proposed a novel NLOS identification method, dubbed the geographic information enhanced (GIE) mechanism, based on \textit{a priori} geographic information and the position estimations of the agents.

Against the above backdrop, we propose a low-cost high-performance geographic and spatio-temporal information based distributed CP (GSTICP) algorithm, for three-dimensional (3D) wireless networks operating in line-of-sight (LOS)/NLOS mixed environments. The main contributions of this paper can be summarized as follows:
\begin{itemize}
\item We provide an FG based information fusion method to fuse the spatio-temporal measurements, and employ the scaled unscented transformation (SUT) \cite{b9} to approximate the nonlinear terms of the messages passing on the FG with high precision and a small number of samples.
\item We exploit the GIE mechanism to identify the NLOS measurements, in which the NLOS identification problem is transformed into a geometric problem by modelling the outlines of buildings as independent rectangles based on their locations in space.
\item We propose an enhanced anchor upgrading mechanism to further decrease the computational complexity of GSTICP by filtering out the agents whose position estimates have already converged so that we can terminate the rest of the iterations.
\end{itemize}

Additionally, our GSTICP supports any type of ranging measurement that can determine the distance between nodes, and it has a much lower computational complexity and communication overhead than the traditional BP based CP  algorithms. Hence, it is more suitable for large-scale heterogeneous wireless networks than the existing CP algorithms. Analysis and simulations results have demonstrated the advantages of our GSTICP.

\textit{Notation}: \(\left\| \cdot \right\|\), \(\rm {E}\{\cdot\}\) and the superscript \([\cdot]^{\text{T}}\) represent the Euclidean norm, expectation and the transpose, respectively; ``\(a\):\(b\)'' denotes the integer-valued time slot vector of [\(a, a+1, \cdots, b\)]; \(p(\cdot)\) represents the probability density function (PDF); \(\delta(\cdot)\) is the Dirac delta function; and \(\sqrt{ \cdot }\) represents the Cholesky decomposition.

\section{System Model and Problem Formulation}
Consider a wireless network composed of \(N\) agents and \(A\) anchors, whose positions are known \textit{a priori} as reference. We assume that the transmission time is slotted, and the agents can move independently from their positions at time slot \(t-1\) to new positions at time slot \(t\). Since our GSTICP is fully distributed, we focus on one agent, i.e., agent \(i\), without loss of generality. In the 3D space, the position of agent \(i\) at time slot \(t\) is denoted by \(\bm{x}_{i}^{t}=\left[x_{i}^{t}, y_{i}^{t}, z_{i}^{t}\right]^{\text{T}}(i=0, \cdots, N+A-1)\). Denote the set of anchors from which agent \(i\) receives signals during time slot \(t\) by \(\mathbb{A}_{\rightarrow i}^{t}\), and the set of agents from which agent \(i\) receives signals during time slot \(t\) by \(\mathbb{U}_{\rightarrow i}^{t}\).

At time slot \(t\), agent \(i\) is able to obtain both the external measurements from its neighbors and the internal measurements based on its own hardware. Denote all the internal measurements and all the external measurements collected by all the agents at time slot \(t\) as the matrix \(\bm{Z}^{t}\). Note that \(\bm{Z}^{t}\) can be broken up into the matrices  \(\bm{Z}_{\text {self}}^{t}\) and \(\bm{Z}_{\text{rel}}^{t}\), where \(\bm{Z}_{\text {self}}^{t}\) consists of all the internal measurements of all the agents, and \(\bm{Z}_{\text{rel}}^{t}\) consists of all the external measurements of all the agents relative to their individual neighbors.

The noise-contaminated ranging measurement from node \(j\) (either agent or anchor)  to agent \(i\) at time slot \(t\) is written as:
\begin{equation}
z_{j \rightarrow i}^{t}=d_{ij}^{t}+e_{j \rightarrow i}+N^{t}_{ij}\label{eq1},
\end{equation}
where \(d_{ij}^{t}\) is the Euclidean distance between agent \(i\) and node \(j\) at time slot \(t\), and it can be measured in a variety of ways, such as time-of-arrival (TOA), angle-of-arrival  (AoA) and received-signal-strength (RSS), to name a few. \(e_{j \rightarrow i}\sim\mathcal{N}\left(0, \sigma_{j \rightarrow i}^{2}\right)\) denotes the measurement error that obeys the Gaussian distribution with zero-mean and variance \(\sigma_{j \rightarrow i}^{2}\), and \(N^{t}_{ij}\) is the NLOS component between agent \(i\) and node \(j\), satisfying:
\begin{equation}
N_{ij}^{t}={\Big\lbrace}\begin{array}{cl}n_{i j}^{t}, & \text {NLOS} \\ 0, & \text {LOS}\end{array} \label{eq2},
\end{equation}
where \(n^{t}_{ij} \sim \mathcal{N}\left(\mu_{ij}^{t}, \sigma_{i j}^{2}\right)\) indicates a Gaussian random variable with mean \(\mu_{ij}^{t}\) and variance \(\sigma_{ij}^{2}\).

The goal of agent \(i\) is to estimate its position \(\bm{x}_{i}^{t}\) at time slot \(t\), given only the external measurements and the internal measurements up to time slot \(t\), i.e., \(p(\bm{x}_{i}^{t} | \bm{Z}^{t})\). We assume that agent \(i\) knows the following information: i) the \textit{a priori} distribution \(p\left(\bm{x}_{i}^{0}\right) \sim \mathcal{N}(\mathrm{E}\{\bm{x}_{i}^{0}\}, \bm{C}_{\bm{x}_{i}^{0}})\) at time slot \(0\), where \(\mathrm{E}\{\bm{x}_{i}^{0}\}=[\mathrm{E}\{x_{i}^{0}\}, \mathrm{E}\{y_{i}^{0}\}, \mathrm{E}\{z_{i}^{0}\}]\) and \(\bm{C}_{\bm{x}_{i}^{0}}=\text{diag}(\sigma_{x_{i}^{0}}^{2}, \sigma_{y_{i}^{0}}^{2}, \sigma_{z_{i}^{0}}^{0})\) represent the expectation and the covariance matrix of \(\bm{x}_{i}^{0}\), respectively; ii) the mobility model characterized by the conditional distribution \(p\left(\bm{x}_{i}^{t} | \bm{x}_{i}^{t-1}\right)\) at any time slot \(t\); iii) the internal measurements \(\bm{z}_{i,\text{self}}^{t}\) at any time slot \(t\) and the corresponding likelihood function \(p\left(\bm{z}_{i,\text{self}}^{t} | \bm{x}_{i}^{t-1}, \bm{x}_{i}^{t}\right)\) at any time slot \(t\); iv) other relevant information obtained from data delivery over the wireless network.

\section{Algorithm Description}
\subsection{Derivation of the SUT Aided Data Fusion Step}
\begin{figure}[t]
	\centering\includegraphics[scale=0.23]{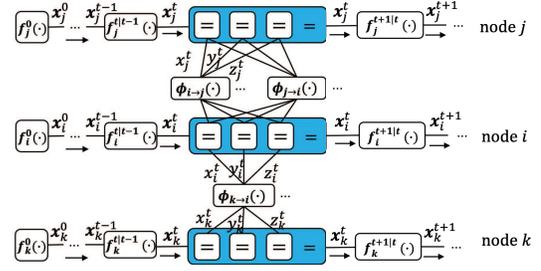}
	\caption{FG of \(p\left(\bm{x}^{0:t} | \bm{Z}^{1:t}\right)\), where nodes \(i, j\in\mathbb{U}_{\rightarrow i}^{t}\), and node \(k\in\mathbb{A}_{\rightarrow i}^{t}\). The arrows represent the temporal flow of the messages passed at different time slots inside a single node (from past to present). We use the following notations: \(f_{i}^{0}(\bm{x}_{i}^{0})=p\left(\bm{x}_{i}^{0}\right)\), \(f_{i}^{t | t-1}(\bm{x}_{i}^{t},\bm{x}_{i}^{t-1})=p\left(\bm{x}_{i}^{t} | \bm{x}_{i}^{t-1}\right) p\left(\bm{z}_{i,\text{self}}^{t} | \bm{x}_{i}^{t-1}, \bm{x}_{i}^{t}\right)\) and \(\phi_{i \rightarrow j}(\bm{x}_{i}^{t}, \bm{x}_{j}^{t})=p\left(z_{i \rightarrow j}^{t} | \bm{x}_{i}^{t}, \bm{x}_{j}^{t}\right)\).}
	\label{fig:1}       
\end{figure}

We first factorize \(p(\bm{x}_{i}^{0:t}|\bm{Z}^{1:t})\) as:
\begin{equation}
\begin{aligned} p(\bm{x}_{i}^{0:t} | \bm{Z}^{1:t}) \ \propto \ & p\left(\bm{x}_{i}^{0}\right) \prod_{\tau=1}^{t}\left\{p\left(\bm{x}_{i}^{\tau} | \bm{x}_{i}^{\tau-1}\right)\right.\\ &\left.\times p\left(\bm{z}_{i,\text{self}}^{\tau} | \bm{x}_{i}^{\tau}, \bm{x}_{i}^{\tau-1}\right) p\left(\bm{Z}_{\text{rel}}^{\tau} | \bm{x}_{i}^{\tau}\right)\right\}, \end{aligned}
\end{equation}
and the Forney-style FG \cite{b10} of \(p(\bm{x}_{i}^{0:t} | \bm{Z}^{1:t})\) has a structure illustrated in Fig. \ref{fig:1}. For each factor, we create a vertex (drawn as a rectangle), and for each variable we create an edge (drawn as a line). When a variable appears in a factor, we connect the edge to the vertex. When a variable appears in more than two factors, an equality vertex is created. For example, the variable \(\bm{x}_{i}^{t}\) appears in factors \(\phi_{i \rightarrow j}(\bm{x}_{i}^{t}, \bm{x}_{j}^{t})\), \(\phi_{j \rightarrow i}(\bm{x}_{i}^{t}, \bm{x}_{j}^{t})\), \(\phi_{k \rightarrow i}(\bm{x}_{i}^{t}, \bm{x}_{k}^{t})\), \(f_{i}^{t | t-1}(\bm{x}_{i}^{t},\bm{x}_{i}^{t-1})\), and \(f_{i}^{t+1 | t}(\bm{x}_{i}^{t},\bm{x}_{i}^{t+1})\), thus we create an equality vertex and label it ``=". Additionally, when agent \(i\) performs external ranging measurement relative to its neighbor node \(j\) at time slot \(t\), we create a factor \(\phi_{j \rightarrow i}(\cdot)\) which is local to agent \(i\) and is a function of the ranging measurement \(z_{j \rightarrow i}^{t}\).

We then execute an iterative SPA on the above FG, as detailed below. The \textit{a posteriori} distribution (usually called \textit{belief}) concerning the \(x\)-component of the position vector of agent \(i\) represents the \(x\)-component of the message broadcast by agent \(i\) at iteration \(l\) and time slot \(t\), i.e., \({b}^{l}(x_{i}^{t})\), and it satisfies:
\begin{equation}
\begin{aligned}
{b}^{l}(x_{i}^{t}) \ \propto \ & \mu_{f_{i}^{(t | t-1)}(\cdot) \rightarrow x_{i}^{t}}(\cdot) \prod_{k \in \mathbb{A}_{\rightarrow i}^{t}} \mu_{\phi_{k \rightarrow i}(\cdot) \rightarrow x_{i}^{t}}^{l}(\cdot) \\& \prod_{j \in \mathbb{U}_{\rightarrow i}^{t}}  \mu_{\phi_{j \rightarrow i}(\cdot) \rightarrow x_{i}^{t}}^{l}(\cdot), \end{aligned}
\end{equation}
where \(\mu_{f_{i}^{(t | t-1)}(\cdot)  \rightarrow x_{i}^{t}}(\cdot)\) is the message from factor \(f_{i}^{(t | t-1)}(\cdot)\) to variable \(x_{i}^{t}\) and it is the \(x\)-component temporal information of agent \(i\) satisfying:
\begin{equation}
\mu_{f_{i}^{(t | t-1)}(\cdot) \rightarrow x_{i}^{t}}(\cdot) \ \propto \ {b}^{l_\text{max}}(x_{i}^{t-1})f_{i}^{t|t-1}(\cdot),
\end{equation}
with \(l_\text{max}\) representing the maximum number of iterations at time slot \(t-1\), while \(\prod_{j \in \mathbb{U}_{\rightarrow i}^{t}}  \mu_{\phi_{j \rightarrow i}(\cdot) \rightarrow x_{i}^{t}}^{l}(\cdot)\) and \(\prod_{k \in \mathbb{A}_{\rightarrow i}^{t}}\mu_{\phi_{k \rightarrow i}(\cdot) \rightarrow x_{i}^{t}}^{l}(\cdot)\) are the spatial ranging information satisfying:
\begin{equation}
\mu_{\phi_{k \rightarrow i}(\cdot) \rightarrow x_{i}^{t}}^{l}(\cdot) \ \propto \ \iiint \phi_{k \rightarrow i}(\cdot) \mu_{x_{k}^{t} \rightarrow \phi_{k \rightarrow i}(\cdot)}^{l}  (\cdot) d \bm{x}_{k}^{t} d y_{i}^{t} d z_{i}^{t},
\end{equation}
and 
\begin{equation}
\mu_{\phi_{j \rightarrow i}(\cdot) \rightarrow x_{i}^{t}}^{l}(\cdot) \ \propto \ \iiint \phi_{j \rightarrow i}(\cdot) \mu_{x_{j}^{t} \rightarrow \phi_{j \rightarrow i}(\cdot)}^{l}  (\cdot) d \bm{x}_{j}^{t} d y_{i}^{t} d z_{i}^{t},
\end{equation}
respectively. In (6), the message from variable \(x_{k}^{t}\) to factor \(\phi_{k \rightarrow i}(\cdot)\), i.e., the belief of node \(k\) broadcast at time slot \(t\) and iteration \(l\) can be written as:
\begin{equation}
\mu_{x_{k}^{t} \rightarrow \phi_{k \rightarrow i}(\cdot)}^{l}(\cdot)=b^{l}(x_{k}^{t})=\delta(x_{k}^{t}-\mathrm{E}\{x_{k}^{t}\}),
\end{equation}
where
\begin{equation}
\begin{aligned}
\begin{array}{c}\phi_{k \rightarrow i}(\cdot) = \frac{1}{\sqrt{2 \pi \sigma_{k \rightarrow i}^{2}}} \exp \left\{-\frac{\left(z_{k \rightarrow i}^{t}-\left\|\bm{x}_{i}^{t}-\bm{x}_{k}^{t}\right\|\right)^{2}}{2 \sigma_{k \rightarrow i}^{2}}\right\},\end{array}
\end{aligned}
\end{equation}
For brevity, we define the exponential function in (9) as \(h_{k}\), where \(k \in \mathbb{A}_{\rightarrow i}^{t} \cup \mathbb{U}_{\rightarrow i}^{t}\). Then substituting (8) and (9) into (6) leads to:
\begin{equation}
\mu_{\phi_{k \rightarrow i}(\cdot) \rightarrow x_{i}^{t}}^{l}(\cdot) \\ \propto \iiint h_{k} d \bm{x}_{k}^{t}d y_{i}^{t}d z_{i}^{t}.
\end{equation}

Unfortunately, (10) involves integrals and it is difficult to obtain close-form expressions for (10) due to the term \(h_{k}\), which is a nonlinear function of \(\bm{x}_{i}^{t}\). As a remedy, we approximate the PDF of \(h_{k}\) by employing the SUT technique, which picks a minimal set of samples (called \textit{sigma points}) around the mean. Then the sigma points are used to replace \(\bm{x}_{i}^{t}\) in the nonlinear term \(h_{k}\) to form the new mean and variance estimates of \(h_{k}\). In general, it is computationally efficient and convenient to generate a symmetric set of 2\(n\)+1 sigma points to define a discrete distribution having a given mean and covariance in \(n\) dimensions \cite{b9}, where we have $n=3$ for the 3D space. First, we draw \(N_{s}\)=2\(n\)+1=7 samples \(\{\bm{v}_{a}\}_{a=0}^{6}\) from the PDF of \(\bm{x}_{i}^{t}\), thus we have:
\begin{equation}
\bm{v}_{0}=\mathrm{E}\{\bm{x}_{i}^{t}\}, \quad a=0;
\end{equation}
\begin{equation}
\bm{v}_{a}=\mathrm{E}\{\bm{x}_{i}^{t}\}+\left(\sqrt{(n+\lambda) \bm{C}_{\bm{x}_{i}^{t}}}\right)_{a},
\end{equation}
when \(a=1, 2, 3\);
\begin{equation}
\bm{v}_{a}=\mathrm{E}\{\bm{x}_{i}^{t}\}-\left(\sqrt{(n+\lambda) \bm{C}_{\bm{x}_{i}^{t}}}\right)_{a-3}, 
\end{equation}
when \(a=4, 5, 6\); and
\begin{equation}
\lambda=3 \left(\alpha^{2}-1\right),
\end{equation}
where \((\cdot)_{a}\) represents the \(a\)th column of a matrix, \(\alpha\) is a small positive number which controls the distribution of the samples. Then, assign a weight to each sample by:
\begin{equation}
w_{0}^{m}=\frac{\lambda}{n+\lambda},
\end{equation}
\begin{equation}
w_{0}^{c}=\frac{\lambda}{n+\lambda}+\left(1-\alpha^{2}+\beta\right),
\end{equation}
\begin{equation}
w_{a}^{m}=w_{a}^{c}=\frac{1}{2(n+\lambda)}, \quad a=1,2, \ldots, 6,
\end{equation}
where \(w_{a}^{m}\) and \(w_{a}^{c}\) denote the weight assigned to the mean and covariance of the \(a\)th sample, respectively; \(\beta\) is a non-negative number and chosen empirically as 2.

Similar to \(\bm{x}_{i}^{t}\) in \(h_{k}\), we perform the same operation on \(\{\bm{v}_{a}\}_{a=0}^{2n}\) and obtain \(\bm{g}\)=\(\{g_{a}\}_{a=0}^{6}\):
\begin{equation}
g_{a}=\exp \left\{-\frac{\left(z_{k \rightarrow i}^{t}-\left\|\bm{v}_{a}-\bm{x}_{k}^{t}\right\|\right)^{2}}{2 \sigma_{k \rightarrow i}^{2}}\right\},
\end{equation}
then the mean and variance of \(h_k\) can be obtained upon using the above SUT, and they satisfy:
\begin{equation}
\mathrm{E}\{h_{k}\}=\sum_{a=0}^{6} w_{a}^{m} g_{a},
\end{equation}
\begin{equation}
\sigma_{h_{k}}^{2}=\sum_{a=0}^{6} w_{a}^{c}{(g_{a}-\mathrm{E}\{h_{k}\})}^{2}. 
\end{equation}

Thus the message \(\mu_{\phi_{k \rightarrow i}(\cdot) \rightarrow x_{i}^{t}}^{l}(\cdot)\) is given by
\begin{equation}
\mu_{\phi_{k \rightarrow i} \rightarrow x_{i}^{t}}^{l}(\cdot) \propto \mathcal{N}(\mathrm{E}\{h_{k}\}, \sigma_{h_{k}}^{2}).
\end{equation}

For brevity, we further denote the message \(\mu_{\phi_{m \rightarrow i}(\cdot) \rightarrow x_{i}^{t}}^{l} (\cdot)\) by \(\mu_{m}^{l}\) (\(m \in \mathbb{A}_{\rightarrow i}^{t} \cup \mathbb{U}_{\rightarrow i}^{t}\)). The above derivation also applies to the message \(\mu_{j}^{l}\):
\begin{equation}
\mu_{\phi_{j \rightarrow i}(\cdot) \rightarrow x_{i}^{t}}^{l} (\cdot)=\mu_{j}^{l} \propto \mathcal{N}\left(\mathrm{E}\{\mu_{j}^{l}\}, (\sigma_{\mu_{j}^{l}})^{2}\right),
\end{equation}
where
\begin{equation}
\mathrm{E}\{\mu_{j}^{l}\}=(\sigma_{\mu_{j}^{l}})^{2}\left(\frac{\mathrm{E}\{h_{j}\}}{(\sigma_{h_{j}})^{2}}+\frac{\mathrm{E}^{l}\{x_{j}^{t}\}}{(\sigma_{x_{j}^{t}}^{l})^{-2}}\right),
\end{equation}
and
\begin{equation}
(\sigma_{\mu_{j}^{l}})^{2}=\left((\sigma_{h_{j}})^{-2}+(\sigma_{x_{j}^{t}}^{l})^{-2}\right)^{-1}.
\end{equation}

Substituting (5), (21) and (22) into (4) we obtain:
\begin{equation}
p(x_{i}^{t}|\bm{Z}^{t})={b}^{l_{\rm {max}}}(x_{i}^{t}) \propto \mathcal{N}(\mathrm{E}\{x_{i}^{t}|\bm{Z}^{t}\}, \sigma_{x_{i}^{t}|\bm{Z}^{t}}^{2}),
\end{equation}
where
\begin{equation}
\begin{aligned}
\mathrm{E}\{x_{i}^{t}|\bm{Z}^{t}\}=&(\sigma_{x_{i}^{t}|\bm{Z}^{t}})^{2}(\frac{\hat{x}_{i}^{t|t-1}}{(\hat{\sigma}_{i,x}^{t|t-1})^{2}}+\sum_{j \in \mathbb{U}_{\rightarrow i}^{t}}\frac{\mathrm{E}\{\mu_{j}^{l_{\text{max}}}\}}{(\sigma_{\mu_{j}^{l_{\text{max}}}})^{2}} \\ & + \sum_{k \in \mathbb{A}_{\rightarrow i}^{t}}\frac{\mathrm{E}\{h_{k}\}}{(\sigma_{h_{k}})^{2}}),\end{aligned}
\end{equation}
and
\begin{equation}
\begin{aligned}
(\sigma_{x_{i}^{t}|\bm{Z}^{t}})^{2}=&(\frac{1}{(\hat{\sigma}_{i,x}^{t|t-1})^{2}}+\sum_{j \in \mathbb{U}_{\rightarrow i}^{t}}\frac{1}{(\sigma_{\mu_{j}^{l_{\text{max}}}})^{2}}\\ & + \sum_{k \in \mathbb{A}_{\rightarrow i}^{t}}\frac{1}{(\sigma_{h_{k}})^{2}})^{-1}.\end{aligned}
\end{equation}

At any time slot \(t\), each agent can determine the minimum mean squared error (MMSE) based estimate of the $x$-component of its own position by taking the mean of \(x_{i}^{t}\), as follows:
\begin{equation}
\hat{x}_{i}^{t}=\int x_{i}^{t} b^{l_{\text{max}}}\left(x_{i}^{t}\right) \mathrm{d} x_{i}^{t},
\end{equation}
while \(\hat{y}_{i}^{t}\) and \(\hat{z}_{i}^{t}\) can be obtained in a similar manner.

\subsection{Geographic Information Based NLOS Identification Mechanism}

In practical implementation, there are dense obstacles distributed in urban areas, and the NLOS measurements in such an environment may severely deviate from the LOS measurements. In this case, the positive bias of NLOS measurements propagates among the whole wireless network via message passing, hence resulting in severe positioning performance degradation. To address this problem, we adopt the scheme of \cite{b8} where the NLOS measurements are identified by a region sampling method based on geographic information and the current node position estimate. First, we obtain the geographic information (e.g., building placement information) from OpenStreetMap \cite{b11}. Then, buildings (even with irregular shapes) are modelled by the minimum possible cubes that best fit the individual size of the buildings. As a result, the NLOS identification is fairly quick, since the cubes can be efficiently stored and quickly retrieved by the searching technique of R-tree. This is important, because the topology of wireless networks can change frequently. 

It is worth noting that if there exists any cube between the position estimates of the two nodes, the measurement between the two nodes is regarded as NLOS. Otherwise, it is regarded as LOS. Additionally, since some buildings may have irregular shapes, the accuracy of NLOS identification can be improved upon using real building shapes. For clarity, our GSTICP is presented in Algorithm \ref{alg1}.

\begin{algorithm} 
	\caption{GSTICP}
	\label{alg1}
	\begin{algorithmic}
	\footnotesize
		\REQUIRE The \textit{a priori} distribution \(p(\bm{x}_{i}^{0})\), the local mobility model \(p(\bm{x}_{i}^{t}|\bm{x}_{i}^{t-1})\), the local likelihood function \(p(\bm{z}_{i,\text{self}}^{t} |\bm{x}_{i}^{t-1}, \bm{x}_{i}^{t})\)
		\ENSURE The \textit{a posteriori} distribution \(\mathrm{E}\{\bm{x}_{i}^{t}|\bm{Z}^{t}\}\) and \(\bm{C}_{\bm{x}_{i}^{t}|\bm{Z}^{t}}\)
		\FOR{node \(i \in \mathbb{U}\)}
		\STATE initialize \(b^{0}(\bm{x}_{i}^{t})= \mu_{f_{i}^{t|t-1}(\cdot)\rightarrow \bm{x}_{i}^{t}}(\cdot)\).
		\STATE collect external measurements.
		\STATE identify the NLOS ranging measurements upon using geographic information. 
		\STATE discard the NLOS ranging measurements.
		\FOR{iteration \(l\) = 1 to \(l_\text{max}\)}
		\STATE broadcast \(b^{l-1}(\bm{x}_{i}^{t})\).
        \STATE receive \(b^{l-1}(\bm{x}_{j}^{t})\).
        \STATE using (4) to calculate the belief.
        \STATE obtain the position information \(\mathrm{E}\{\bm{x}_{i}^{t}|\bm{Z}^{t}\}\) and \(\bm{C}_{\bm{x}_{i}^{t}|\bm{Z}^{t}}\) by (26) and (27).
        \STATE using (28) to estimate the position of node $i$.
		\ENDFOR 
		\ENDFOR
	\end{algorithmic} 
\end{algorithm}

\subsection{Acceleration of GSTICP}
We conceive an enhanced anchor upgrading (EAU)  technique to further decrease the computational complexity of the proposed GSTICP algorithm. The EAU consists of anchor upgrading and iteration reduction. We set two positive thresholds \(\eta_\text{1}\) and \(\eta_\text{2}\), satisfying \(\eta_\text{1}<\eta_\text{2}\), for \(\mathrm{E}\{x_{i}^{t}|\bm{Z}^{t}\}\).

\subsubsection{Anchor upgrading}
In time slot \(t\), after several iterations, the individual position estimates of some agents become converge and are almost unchanged in the subsequent iterations. If \(\mathrm{E}\{x_{i}^{t}|\bm{Z}^{t}\}\) at two consecutive iterations changes less than the threshold \(\eta_\text{1}\), then agent \(i\) will be upgraded as a pseudo-anchor, and the \textit{belief} concerning the position  of agent \(i\) will not be updated within this time slot.

\subsubsection{Iteration reduction}
If \(\mathrm{E}\{x_{i}^{t}|\bm{Z}^{t}\}\) at two consecutive iterations changes less than the threshold \(\eta_\text{2}\) but larger than the threshold \(\eta_\text{1}\), then the update of the \textit{belief} will skip at the next iteration. More specifically, if 
\begin{equation}
|\mathrm{E}^{l+1}\{x_{i}^{t}|\bm{Z}^{t}\}-\mathrm{E}^{l}\{x_{i}^{t}|\bm{Z}^{t}\}|<\eta_\text{1}
\end{equation}
is satisfied at time slot \(t\), then set \({b}^{l+}(x_{i}^{t})={b}^{l}(x_{i}^{t}), l_{+}=l+2,...,l_\text{max}\). The ranging measurements among the  pseudo-anchors as well as the ranging measurements between a pseudo-anchor and an anchor are also deleted. Moreover, if
\begin{equation}
\eta_\text{1}<|\mathrm{E}^{l+1}\{x_{i}^{t}|\bm{Z}^{t}\}-\mathrm{E}^{l}\{x_{i}^{t}|\bm{Z}^{t}\}|<\eta_\text{2},
\end{equation}
then set \({b}^{l+2}(x_{i}^{t})={b}^{l+1}(x_{i}^{t})\). Therefore, the number of samples are decreased and the computational complexity is substantially reduced, despite at the cost of possibly marginally increased positioning error, as demonstrated by the subsequent analysis and simulation results.

\begin{table*}[t]
  \scriptsize
  \caption{Comparison of the computational complexity, the communication overhead, and the number of samples (\(N_\text{s}\))}
  \label{tab:1}       
  \centering
   \begin{tabular}{|c|c|c|c|} 
    \hline 
    Algorithm&Complexity&Communication overhead&\(N_\text{s}\)\\
    \hline  
    GSTICP&\(\mathcal{O}(N_\text{rel} \cdot l_\text{max} + N_\text{rel} \cdot l_\text{max} \cdot \left(\text{log}_{2}N_\text{building} + \text{log}_{2}N_\text{rel} \right))\)&14&7\\
    \hline
    SPA-TE with GIE mechanism&\(\mathcal{O}(N_\text{rel} \cdot l_\text{max} + N_\text{rel} \cdot l_\text{max}\cdot \left(\text{log}_{2}N_\text{building} + \text{log}_{2}N_\text{rel} \right))\)&2&N/A\\
    \hline
    GIE-UCL&\(\mathcal{O}\left(N_\text{s} \cdot N_\text{rel} \cdot l_\text{max} + N_\text{rel} \cdot l_\text{max} \cdot \left(\text{log}_{2}N_\text{building} + \text{log}_{2}N_\text{rel} \right) \right)\)&\(\mathcal{O}(N_\text{rel})\)&Large \\
    \hline
    SPAWN (without NLOS-link identification) &\(\mathcal{O}(N_\text{s}^{2} \cdot N_\text{rel})+\mathcal{O}(N_\text{s})\)&\(\mathcal{O}(N_\text{s})\)&Large\\
    \hline
    \end{tabular}%
\end{table*}%

\subsection{Computational Complexity and Communication Overhead}
We compare our GSTICP, the SPA-TE \cite{b5} with the GIE mechanism, GIE-UCL \cite{b6}, and SPAWN \cite{b4} (without NLOS-link identification) in terms of the computational complexity, the communication overhead, and the number of samples in Table \ref{tab:1}, where \(N_\text{rel}\) denotes the number of agent's neighbor nodes, \(N_\text{s}\) is the number of samples, and \(N_\text{building}\) represents the number of buildings in the area of interest.   Since the above algorithms are fully distributed, we only evaluate the computational complexity and communication overhead of a single agent, and \(N_\text{s}\) is evaluated in terms of one message. Furthermore, since all the algorithms considered share common procedures in ranging measurement and neighbor discovery, it is sufficient to only compare their computational complexity and their communication overhead caused by the information fusion step and the NLOS-link identification step. For the proposed GSTICP, the complexity of the information fusion step is related to the number of neighbor nodes \(N_\text{rel}\) and the number of iterations \(l_\text{max}\), whereas the complexity of the NLOS identification is on the order of \(\mathcal{O}(N_\text{rel} \cdot \left(\text{log}_{2}N_\text{building} + \text{log}_{2}N_\text{rel} \right))\). Since all the messages are broadcast and updated relying on the Gaussian distribution assumption, only the approximate mean and the covariance of the seven sigma points need to be broadcast per iteration. As a result, the communication overhead of GSTICP is 14. Additionally, the number of samples that come from the SUT step used in GSTICP is seven, while SPAWN and GIE-UCL rely on a large number of samples.

Besides, we have the following observations. First of all, the computational complexity of our GSTICP is much less than that of GIE-UCL, and the information fusion complexity of GSTICP is much less than that of SPAWN. This should be expected since GIE-UCL and SPAWN utilized thousands of samples to approximate the nonlinear messages passing on the FG, whereas our GSTICP only exploited seven sigma points. Secondly, the communication overhead of our GSTICP is marginally higher than that of SPA-TE with GIE. This coincides with our intuition, since SPA-TE only broadcasts the mean  and covariance of the messages approximated by parameterized method, whereas our GSTICP broadcasts the mean and covariance of seven sigma points. However, this minor extra cost is tolerable, because GSTICP is able to provide a much more accurate location information.

\section{Simulation Results and Discussions}
\begin{figure}[tbp]
	\centering\includegraphics[scale=0.24]{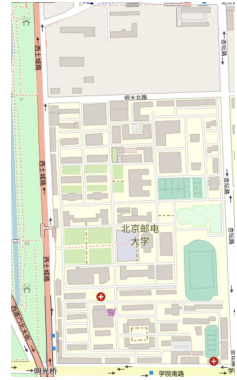}
	\caption{Map of our university, obtained from OpenStreetMap.}
	\label{fig:2}       
\end{figure}

We evaluate the performance of our GSTICP algorithm against several representative CP algorithms in numerical simulations. The positioning performance is evaluated by the cumulative distribution function (CDF) \(P(e \leq \epsilon [\text{m}])\), where \(e\) denotes the positioning error and is characterized by the mean squared error between the estimated positions and the true position, and \(\epsilon\) represents the allowed positioning error. We employ our university as the interested area, as shown in Fig.~\ref{fig:2}. Consider a wireless network with 80 agents and 15 anchors, which are randomly scattered into the above \((1000 \times 600 \times 50) \text{m}^{3}\) interested space. The distance within which to perform ranging and communication is set to 300\(\text{m}\). The default number of iterations \(l_\text{max}\) is set to 20. Each agent moves a distance \(d_{i}^{t}\) in a random direction, with \(d_{i}^{t}\sim\mathcal{N}(0,1)\).
\begin{figure}[t]
	\centering\includegraphics[scale=0.4]{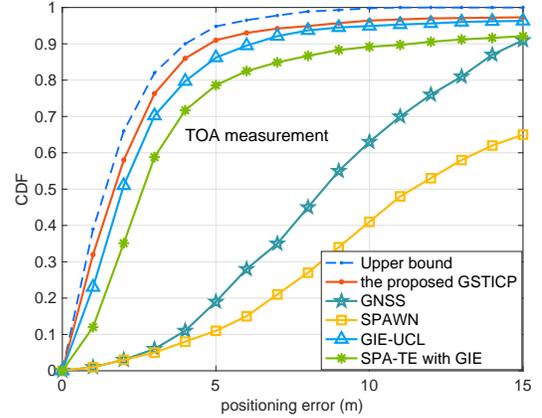}
	\caption{The CDFs of the GSTICP, SPAWN, GIE-UCL, SPA-TE with GIE under TOA measurements and of the GNSS based method. The “Upper Bound” is the simulation results of GSTICP with known LOS/NLOS measurements.}
	\label{fig:3}       
\end{figure}
\vspace{-0.5cm}

We first compare the performance of our GSTICP against the SPAWN, SPA-TE with GIE, and GIE-UCL schemes under TOA measurements, as well as against the GNSS based method, in Fig. \ref{fig:3}. In addition, the simulation results of GSTICP with known LOS/NLOS measurements are provided as the “Upper Bound”. We have the following observations. First of all, GSTICP outperforms the SPAWN, SPA-TE with GIE, and GIE-UCL schemes. For example, \(P\left(e \leq 4[\text{m}]\right)\) is 0.86 for GSTICP, while it is 0.80, 0.72 and 0.09 for the GIE-UCL, SPA-TE with GIE, and SPAWN schemes, respectively. This is attributed to the gains from internal temporal information and the higher accuracy of approximation of SUT adopted in our GSTICP. Secondly, SPA-TE with GIE performs worse than GIE-UCL. This finding is consistent with our intuition that the approximation accuracy by using the first-order Taylor expansion to replace nonlinear terms is lower than that of using a large number of samples. Thirdly, SPAWN performs even worse than the GNSS based method, since it utilizes NLOS measurements, whose endogenous bias is propagated over the wireless network and causes performance degradation. This indicates that the NLOS identification is crucial to wireless positioning in practical wireless networks.
\begin{figure}[t]
	\centering\includegraphics[scale=0.4]{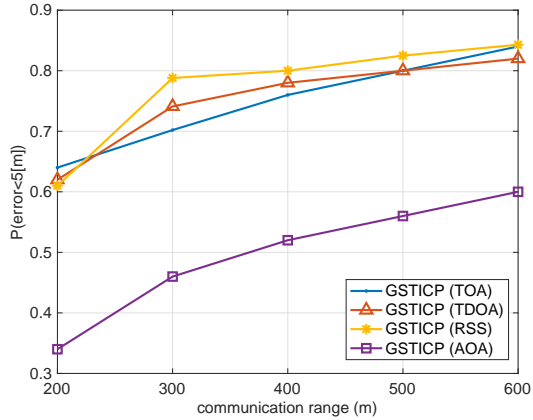}
	\caption{CDFs of different measurements versus the communication range.}
	\label{fig:4}       
\end{figure}
\begin{figure}[t]
	\centering\includegraphics[scale=0.4]{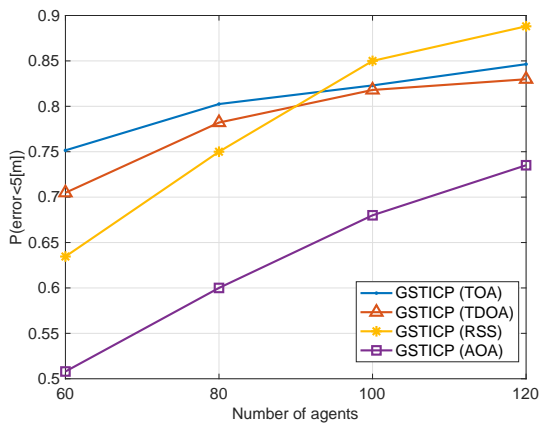}
	\caption{CDFs of different measurements versus the agent density.}
	\label{fig:5}       
\end{figure}

Fig. \ref{fig:4} and Fig. \ref{fig:5} illustrate the impact imposed by the communication range and the agent density on the positioning performance, respectively, under the positioning accuracy requirement of \(P\left(e\leq 5 [\text{m}]\right)\). In Fig. \ref{fig:4}, the number of agents and anchors is set to 80 and 20, respectively, whereas the communication range is increased from 200m to 600m. By contrast, in Fig. \ref{fig:5}, the communication range is set to 400m and the number of anchors remains 20, whereas the total number of agents is increased from 60 to 120. The results are compliant with our intuition that the positioning performance could be improved by increasing the communication range and the agent density.

\section{Conclusion}
We have proposed a low-complexity GSTICP algorithm for wireless localization that supports various types of ranging measurements under the LOS/NLOS mixed environments. To reduce the computational complexity and communication overhead, we first utilized the symmetric sampling based SUT technique to approximate the nonlinear terms of the messages with high approximation accuracy and a dramatically small number of sample points. Moreover, we proposed the EAU mechanism to filter out the agents whose position estimates have already converged so that we can terminate the rest of the iterations. In order to identify the NLOS measurements, a scheme with \textit{a priori}  geographic information was exploited based on the R-tree searching method. Analysis and simulation results validated that our GSTICP has achieved competitive positioning performance, at the cost of a much lower computational complexity than the traditional sample based CP algorithms in the LOS/NLOS mixed environments. Moreover, we demonstrated that the positioning accuracy of GSTICP can be improved by increasing the communication range and the agent density.

\end{document}